\documentclass[12pt,a4paper,english]{revtex4}
\usepackage{babel}

\usepackage{amsmath}
\usepackage{amssymb}
\usepackage{amsthm} 

\usepackage{array}

\usepackage{slashed}

\usepackage[dvips]{graphicx} 
\usepackage{subfigure}

\usepackage{varioref}   
\usepackage{xr}         

\usepackage{pstricks}           

\allowdisplaybreaks[1]

\newcommand\nn{\nonumber}
\newcommand\ba{\begin{eqnarray}}
\newcommand\ea{\end{eqnarray}}
\newcommand\eq[1] {\begin{align} #1 \end{align}}   
\newcommand\ga[1] {\begin{gather} #1 \end{gather}}   

\newcommand{\br}[1]{\left( #1 \right)}
\newcommand{\brs}[1]{\left[ #1 \right]}

\newcommand{\brm}[1]{\left| #1 \right|}

\newcommand{\Li}[2]{\mbox{Li}_{#1}\br{#2}}
\newcommand{\Sp}{\mbox{Sp}}

\renewcommand{\Re}{\mbox{Re}}

\newcommand{\vv}[1]{{\bf #1}}

\newcommand{\dd}[1]{{\slashed #1}}   

\renewcommand{\P} {{\mathcal P}}

\newcommand{\Moller} {{M{\o}ller}~} 

\begin{document}

\title{One-loop chiral amplitudes of \Moller scattering process}

\author{A.~I.~Ahmadov}
\email{ahmadov@theor.jinr.ru}
\affiliation{Institute of Physics, Azerbaijan National Academy of Science, Baku, Azerbaijan}
\affiliation{Joint Institute for Nuclear Research, Dubna, Russia}

\author{Yu.~M.~Bystritskiy}
\email{bystr@theor.jinr.ru}
\affiliation{Joint Institute for Nuclear Research, Dubna, Russia}

\author{A.~N.~Ilyichev}
\email{ily@hep.by}
\affiliation{National Center of Particle and High Energy Physics of Belarussian State University, Minsk, Belarus}

\author{E.~A.~Kuraev}
\email{kuraev@theor.jinr.ru}
\affiliation{Joint Institute for Nuclear Research, Dubna, Russia}

\author{V.~A.~Zykunov}
\email{vladimir.zykunov@cern.ch}
\affiliation{Belarussian State University of Transport, Gomel, Belarus}

\begin{abstract}
The high energy amplitudes of the large angles \Moller  scattering are calculated in
frame of chiral basis in Born and 1-loop QED level.
Taking into account as well the
contribution from emission of soft real photons the compact relations free from
infrared divergences are obtained. The expressions for separate chiral amplitudes contribution to
the cross section are in agreement with renormalization group predictions.
\end{abstract}

\maketitle

\section{Introduction}

Main prediction of Standard Model (SM) --- Parity Violation (PV) --- can be experimentally measured
in \Moller scattering process (quasi-elastic electron-electron scattering)
due to possible contributions from mechanism with massive SM boson exchange between the electrons.
The first PV observation in M{\o}ller scattering was made by E-158 experiment
at SLAC \cite{Kumar:1995ym,Kumar:2007zze,Anthony:2003ub}, which studied scattering of 45- to 48-GeV polarized electrons on the
unpolarized electrons of a hydrogen target.
Its result at low $Q^2$ = 0.026 $\mbox{GeV}^2$ for PV asymmetry
$A_{LR} = (1.31 \pm 0.14\ \mbox{(stat.)} \pm 0.10\ \mbox{(syst.)}) \times 10^{-7}$ \cite{Anthony:2005pm}
allowed one
of the most important parameters in the Standard Model -- the sine of the  Weinberg angle --
to be determined with unprecedented accuracy.
The next-generation experiment to study $e-e$ scattering, MOLLER \cite{Benesch:2011},
planned at JLab following the 11 GeV upgrade,
will offer a new level of sensitivity and measure the PV asymmetry in the scattering
of longitudinally polarized electrons off unpolarized target to a precision of 0.73 ppb.
That would allow a determination of the weak mixing angle with an uncertainty of about 0.1\%, a factor of five
improvement in fractional precision over the measurement by E-158.

The polarized M{\o}ller scattering
has allowed the high-precision determination of the electron-beam
polarization at many experimental programs, such as
SLC \cite{Swartz:1994yv}, SLAC \cite{Steiner:1998gf,Band:1997ee}, JLab \cite{Hauger:1999iv} and MIT-Bates \cite{Arrington:1992xu}.
A M{\o}ller polarimeter may also be useful in future experiments planned at the ILC \cite{Alexander:2000bu}.

Since the polarized M{\o}ller scattering is a very clean process with a well-known
initial energy and  extremely suppressed backgrounds, any inconsistency with the Standard Model
will signal new physics.
Some examples of physics beyond the Standard Model to which M{\o}ller scattering measurement extends sensitivity
include new neutral bosons ($Z'$), electron compositeness, supersymmetry and doubly charged scalars  \cite{Benesch:2011}.
Thus M{\o}ller scattering experiments can provide indirect access to physics
at multi-TeV scales and play an important complementary role to the LHC research program \cite{Heusch:2000pj,Feng:1999zv}.

Moreover the accuracy of calculations in Born approximation (sometimes even on one-loop level)
seems to be insufficient to compare with the results of modern experiments.
A significant theoretical effort has been dedicated to one-loop radiative corrections already.
A short review of the literature to date on that topic is done in \cite{Aleksejevs:2010ub}.
The motivation of our paper is to calculate radiative corrections
(on this stage within the QED framework but planning in the future to add weak contribution)
in one-loop approximation, in order to show up the correct renormalization group behavior of
leading logarithms contribution.
This will allow us to extend the applicability of this calculation to higher order corrections.

The paper is organized as follows.
After a brief review of results in Born approximation and formalism of chiral
amplitudes in Section~\ref{SectionBorn}, the virtual corrections associated with the known expressions of vertex
functions as well as ones arising from polarization of vacuum are considered in Section~\ref{SectionVertexSoft}.
As well the radiative corrections from emission of real soft photon are presented there.
In Section~\ref{SectionBox} the contribution from box type Feynman diagrams are considered.
In Section~\ref{SectionRenormalizationGroup} the total result for the chiral amplitudes and
the relevant contributions to the cross sections are  presented.
In Section~\ref{SectionConclusion} we make some concluding remarks about the accuracy of our result.
Finally, in Appendix~\ref{appendix.LoopIntegrals} we present some useful integrals for box-diagrams calculation.

\section{Born level chiral amplitudes of Moller scattering process}
\label{SectionBorn}

We consider the process of electron--electron scattering (so called \Moller process \cite{Moeller:1932})
\eq{
    e^-\br{p_1,\lambda_1} + e^-\br{p_2,\lambda_2} \to e^-\br{p_1',\lambda_1'} + e^-\br{p_2',\lambda_2'},
    \label{eq:MollerScattering}
}
where $\lambda_{1,2}=\pm$ and $\lambda_{1,2}'=\pm$ are the chiral states of initial and final particles.
Matrix element of this process within the QED has two terms
(see Fig.~\ref{Fig:BornApproximation}):
\begin{figure}
    \centering
    \mbox{
        \subfigure[]{\includegraphics[width=0.2\textwidth]{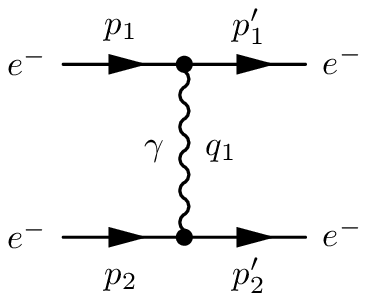}\label{FigBornGamma1}}
        \quad
        \subfigure[]{\includegraphics[width=0.2\textwidth]{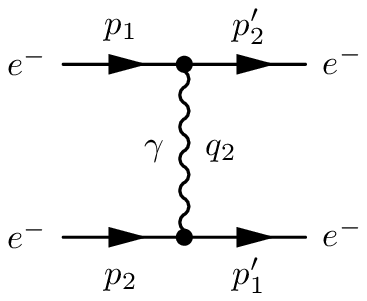}\label{FigBornGamma2}}
    }
    \caption{Born approximation diagrams.}
    \label{Fig:BornApproximation}
\end{figure}
\eq{
    M^{(0)} = M^{(0)}_{\gamma_1} - M^{(0)}_{\gamma_2}
,
}
where $M^{(0)}_{\gamma_1}$ is the contribution which goes from photon exchange in $t$-channel
(Fig.~\ref{FigBornGamma1}),
$M^{(0)}_{\gamma_2}$ is also the contribution of the photon in intermediate state but with the
interchange of final leptons ($p_1' \leftrightarrow p_2'$) which should be done in order to take into account the Pauli principle (see Fig.~\ref{FigBornGamma2}).
The explicit form of these terms are the following
(we use the t'Hooft--Feynman gauge, i.e. $\xi=1$):
\eq{
    M^{(0)\lambda_1\lambda_2\lambda_1'\lambda_2'}_{\gamma_1} &= \frac{e^2}{q_1^2}
    \brs{\bar u^{\br{\lambda_1'}}\br{p_1'} \gamma^\mu u^{\br{\lambda_1}}\br{p_1}}\brs{\bar u^{\br{\lambda_2'}}\br{p_2'} \gamma_\mu u^{\br{\lambda_2}}\br{p_2}}, \nn\\
    M^{(0)\lambda_1\lambda_2\lambda_1'\lambda_2'}_{\gamma_2} &= \frac{e^2}{q_2^2}
    \brs{\bar u^{\br{\lambda_2'}}\br{p_2'} \gamma^\mu u^{\br{\lambda_1}}\br{p_1}}\brs{\bar u^{\br{\lambda_1'}}\br{p_1'} \gamma_\mu u^{\br{\lambda_2}}\br{p_2}}, \nn
}
where $q_1 = p_1 - p_1' = p_2' - p_2$ and $q_2 = p_1 - p_2' = p_1' - p_2$ are the transferred momenta in direct and exchanged diagram,
$e$ is the positron electric charge.

We will consider the process using chiral amplitudes approach. In this approach you select the specific chiral
spin state of initial and final particles:
\eq{
    u^{\br{\lambda}} = \omega_\lambda u,
    \qquad
    \bar u^{\br{\lambda}} = \bar u \, \omega_{-\lambda},
    \qquad
    \lambda = \pm 1 = R,L,
}
where projective operators $\omega_\lambda$ has a form:
\eq{
    \omega_\lambda = \frac{1}{2}\br{1+\lambda\gamma_5},
    \qquad
    \omega_\lambda^2 = \omega_\lambda,
    \qquad
    \omega_+ \omega_- = 0,
    \qquad
    \omega_+ + \omega_- = 1.
}
For the case of massless fermions the completeness condition take place
\eq{
    u^{\br{\lambda}}\br{p}\bar{u}^{\br{\lambda}}\br{p}=\omega_\lambda \dd{p}.
}
Keeping in mind the parity conservation in QED $M^{\lambda}=M^{-\lambda}$ we have for the
summed on polarization matrix element square
\ba
\sum_{spins}|M|^2=2[|M^{++++}|^2+|M^{+-+-}|^2+|M^{+--+}|^2].
\ea

Let us calculate the QED contribution of $RR \to RR$ chiral amplitude, with all fermions of right hand ones $u^R=\omega_+u$:
\eq{
    M^{(0)++++}_{\gamma_1} &= \frac{e^2}{t}
    \brs{\bar u\br{p_1'} \omega_- \gamma^\mu \omega _+ u\br{p_1}}\brs{\bar u\br{p_2'} \omega_- \gamma_\mu \omega_+ u\br{p_2}}, \nn
    \\
    M^{(0)++++}_{\gamma_2} &= \frac{e^2}{u}
    \brs{\bar u\br{p_2'} \omega_- \gamma^\mu \omega _+ u\br{p_1}}\brs{\bar u\br{p_1'} \omega_- \gamma_\mu \omega_+ u\br{p_2}}, \nn
}
where we have used the kinematics Mandelstam invariants in electron mass vanishing limit ($m_e \to 0$):
\eq{
s &= \br{p_1+p_2}^2 = 2 \br{p_1 p_2} = 2 \br{p_1' p_2'}, \nn\\
t &= \br{p_1-p_1'}^2 = -2 \br{p_1 p_1'} = -2 \br{p_2 p_2'}, \\
u &= \br{p_1-p_2'}^2 = -2 \br{p_1 p_2'} = -2 \br{p_2 p_1'}, \nn\\
&s+t+u = 0.\nn
}
In order to transform this amplitudes into calculable traces we multiply the terms which contain factor $\br{1/t}$ by the following quantity:
\eq{
    \frac{a \, b}{a \, b}= 1, \label{EqFactorAB}
}
and the terms with factor $\br{1/u}$ by
\eq{
    \frac{c \, d}{c \, d}= 1,\label{EqFactorCD}
}
where
\eq{
    a &= \bar u\br{p_1} \omega_- \dd{p_2} \omega_+ u\br{p_2'},  &c &= \bar u\br{p_1} \omega_- \dd{p_2} \omega_+ u\br{p_1'}, \label{EqABCD}\\
    b &= \bar u\br{p_2} \omega_- \dd{p_1} \omega_+ u\br{p_1'},  &d &= \bar u\br{p_2} \omega_- \dd{p_1} \omega_+ u\br{p_2'}. \nn
}
After this we obtain trace in numerator and can calculate it immediately:
\eq{
    M^{(0)++++}_{\gamma_1} &= \frac{e^2}{t}
    \frac{1}{a\, b}
    \Sp\brs{\dd{p_1'}\gamma_\mu \omega_+ \dd{p_1} \dd{p_2} \omega_+ \dd{p_2'} \gamma^\mu \omega_+ \dd{p_2} \dd{p_1} \omega_+}
    =
    \frac{e^2}{t}
    \frac{1}{a \, b}
    2 s^2 t, \\
    M^{(0)++++}_{\gamma_2} &= \frac{e^2}{u}
    \frac{1}{c\, d}
    \Sp\brs{\dd{p_2'}\gamma_\mu \omega_+ \dd{p_1} \dd{p_2} \omega_+ \dd{p_1'} \gamma^\mu \omega_+ \dd{p_2} \dd{p_1} \omega_+}
    =
    \frac{e^2}{u}
    \frac{1}{c \, d}
    2 s^2 u. \nn
}
Thus QED amplitude in Born approximation have a form:
\eq{
    M^{(0)++++}_{\gamma} &= M^{(0)++++}_{\gamma_1} - M^{(0)++++}_{\gamma_2} =
    2\br{4\pi\alpha}i s^2 M^0_\gamma,
    \qquad
    M^0_\gamma = \frac{1}{a \, b} - \frac{1}{c \, d};
\nn
    \\
    M^{(0)+-+-}_{\gamma} &= 2(4\pi\alpha)i\frac{u^2}{t c_1d_1}, \\
    M^{(0)+--+}_{\gamma} &= -2(4\pi\alpha)i\frac{t^2}{u c_2d_2}, \nn
}
where $c_1$ and $d_1$ are the modified factors similar to (\ref{EqFactorAB}) or (\ref{EqFactorCD}) which in this case have a form:
\eq{
    c_1 &= \bar u\br{p_1} \omega_- u\br{p_2'},  &d_1 &= \bar u\br{p_2} \omega_+ u\br{p_1'}, \nn \\
    c_2 &= \bar u\br{p_1} \omega_- u\br{p_1'},  &d_2 &= \bar u\br{p_2} \omega_+ u\br{p_2'}. \label{EqC1D1}
}
Using the relations
\ga{
|a|^2=|b|^2=-st, \qquad |c|^2=|d|^2=-su, \qquad a b c^* d^* = -s^2 t u, \nn \\
|c_1|^2=|d_1|^2=-u, \qquad |c_2|^2=|d_2|^2=-t, \nn
}
we obtain for the sum of squares of all six amplitudes the known result \cite{Baier:1973,Akhiezer:1981}
\eq{
    \sum_{\br{\lambda}}\brm{M_{\gamma}^{\br{0}\lambda}}^2
    &=
    8\br{4\pi\alpha}^2
    \brs{
        \br{\frac{s^2}{t^2}+\frac{s^2}{u^2}+\frac{2s^2}{tu}}+
        \br{\frac{t^2}{u^2}+\frac{u^2}{t^2}}
    }=
    8\br{4\pi\alpha}^2\frac{s^4+t^4+u^4}{t^2u^2}.
}
\section{Vertex, polarization of vacuum and soft real photon emission contributions}
\label{SectionVertexSoft}

Born level and virtual photons emission radiative corrections to chiral amplitudes can be put in form
\eq{
    M^\lambda=M^{(0)\lambda}+M^\lambda_{V\Pi B},
}
where $\lambda$ denotes different chiral states.
In this section we will consider the contribution from Feynman diagrams of vertex and photon vacuum polarization types
(see Fig.~\ref{FigVertex},~\ref{FigVP}):
\begin{figure}
    \centering
    \mbox{
        \subfigure[]{\includegraphics[width=0.2\textwidth]{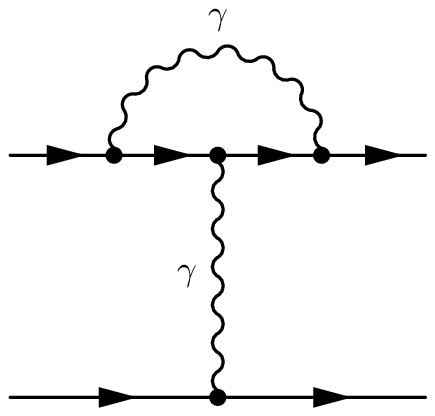}\label{FigVertex}}
        \quad
        \subfigure[]{\includegraphics[width=0.2\textwidth]{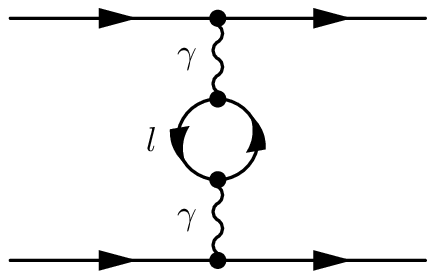}\label{FigVP}}
    }
    \caption{Diagrams of vacuum polarization and vertex correction types.}
    \label{fig.VPandVertex}
\end{figure}
\eq{
    M^{++++}&=8\pi\alpha i s^2\brs{\frac{1}{ab}\br{1+\Pi_t+2\Gamma_t}-\frac{1}{cd}\br{1+\Pi_u+2\Gamma_u}}; \nn \\
    M^{+-+-}&=\frac{8\pi\alpha i u^2}{tc_1d_1}\br{1+\Pi_t+2\Gamma_t};  \\
    M^{+--+}&=\frac{8\pi\alpha i t^2}{uc_2d_2}\br{1+\Pi_u+2\Gamma_u},\nn
}
where vertex and vacuum polarization operators are well known (see \cite{Akhiezer:1981}):
\eq{
    \Gamma_t&=\frac{\alpha}{2\pi}\brs{-l_\lambda\br{l_t-1}+\frac{3}{2}l_t-\frac{1}{2}l_t^2-2+\zeta_2},\qquad \zeta_2=\frac{\pi^2}{6},\nn \\
    \Pi_t&=\frac{\alpha}{3\pi}\br{l_t-\frac{5}{3}}, \nn \\
    \Gamma_u&=\Gamma_t\br{t\to u},\qquad \Pi_u=\Pi_t\br{t\to u}, \qquad l_t=\ln\frac{-t}{m^2}, \qquad l_\lambda=\ln\frac{m^2}{\lambda^2},\nn
}
and  $\lambda$ is a fictitious "photon mass".

Corrections to chiral amplitudes squared from polarization operators are
\eq{
    \Delta_\Pi|M^{++++}|^2 &= -2(8\pi\alpha)^2s^3 \br{\frac{1}{t^2u}\Pi_t+\frac{1}{u^2t}\Pi_u}, \nn \\
    \Delta_\Pi|M^{+-+-}|^2 &= 2(8\pi\alpha)^2\frac{u^2}{t^2}\Pi_t;  \\
    \Delta_\Pi|M^{+--+}|^2 &= 2(8\pi\alpha)^2\frac{t^2}{u^2}\Pi_u. \nn
}
And the similar expression for the vertex operators corrections
\eq{
    \Delta_V|M^{++++}|^2 &= -4(8\pi\alpha)^2s^3\br{\frac{1}{t^2u}\Gamma_t+\frac{1}{u^2t}\Gamma_u}, \nn \\
    \Delta_V|M^{+-+-}|^2 &= 4(8\pi\alpha)^2\frac{u^2}{t^2}\Gamma_t; \nn \\
    \Delta_V|M^{+--+}|^2 &= 2(8\pi\alpha)^2\frac{t^2}{u^2}\Gamma_u.
}
Soft real photons emission  contribution to the squares of chiral amplitudes are \cite{Akhiezer:1981}:
\eq{
    |M^{\lambda}_{soft}|^2&=\delta_{soft}|M^{(0)\lambda}|^2, \nn \\
    \delta_{soft}&=-\frac{4\pi\alpha}{16\pi^3}\int'\frac{d^3 k}{\omega}\br{\frac{p_1}{p_1k}+\frac{p_2}{p_2k}-\frac{p'_1}{p'_1k}-\frac{p'_2}{p'_2k}}^2,
}
where $\int'$ means that photon energy $\omega=\sqrt{\vv{k}^2+\lambda^2}$ do not exceed some small value $\omega<\Delta E \ll E=\sqrt{s}/2$,
where $\Delta E$ and $E$ are defined in center of mass system.
Standard calculation \cite{'tHooft:1978xw} leads to
\ga{
    \delta_{soft}=\frac{2\alpha}{\pi}\brs{\br{l_s-1+L}
\br{l_\lambda+2 l_\epsilon}+\frac{1}{2}l_s^2+Ll_s+K_{soft}}, \nn \\
    l_s=\ln\frac{s}{m^2}, \qquad l_\epsilon=\ln\frac{\Delta E}{E}, \qquad L=l_{ts}+l_{us}, \qquad l_{ts}=\ln\frac{-t}{s},
\qquad l_{us}=\ln\frac{-u}{s}, \nn \\
    K_{soft}=\frac{1}{2}l_{ts}^2+\frac{1}{2}l_{us}^2-2\xi_2+\Li{2}{\cos^2\br{\theta/2}}+\Li{2}{\sin^2\br{\theta/2}},
}
and $\theta$ is the angle (center of mass frame implied) between the directions of motion of one of initial
and the scattered electrons.

Combining the vertex and soft parts of the corrections we obtain
\eq{
    \Delta_{VS}|M^{++++}|^2&=(8\pi\alpha)^2\frac{2\alpha}{\pi}
    \left[
        \frac{s^4}{t^2u^2}\brs{\br{l_s-1}\br{2l_\epsilon+\frac{3}{2}}+2l_\epsilon L+\frac{3}{2}+K_{soft}}+
    \right.\nn \\
    &+\left.
        \frac{s^3}{t^2u}\brs{\frac{1}{2}l_{ts}^2-\frac{3}{2}l_{ts}+2-\zeta_2}+\frac{s^3}{u^2t}\brs{\frac{1}{2}l_{us}^2-\frac{3}{2}l_{us}+2-\zeta_2}
    \right] \nn \\
    \Delta_{VS}|M^{+-+-}|^2&=(8\pi\alpha)^2\frac{2\alpha}{\pi}\frac{u^2}{t^2}
    \left[
        \br{l_s-1}\br{2l_\epsilon+\frac{3}{2}}+2l_\epsilon L+l_{us}\br{l_\lambda+l_s}
    \right. \nn \\
    &-\left.
        \frac{1}{2}+K_{soft}+\frac{3}{2}l_{ts}-\frac{1}{2}l_{ts}^2+\zeta_2
    \right]; \label{eq.SoftVirt} \\
    \Delta_{VS}|M^{+--+}|^2&=(8\pi\alpha)^2\frac{2\alpha}{\pi}\frac{t^2}{u^2}
    \left[
        \br{l_s-1}\br{2l_\epsilon+\frac{3}{2}}+2l_\epsilon L+l_{ts}\br{l_\lambda+l_s}
    \right. \nn \\
    &-\left.
        \frac{1}{2}+K_{soft}+\frac{3}{2}l_{us}-\frac{1}{2}l_{us}^2+\zeta_2
    \right].\nn
}

\section{Box diagrams contribution}
\label{SectionBox}

\begin{figure}
    \centering
    \mbox{
        \subfigure[]{\includegraphics[width=0.2\textwidth]{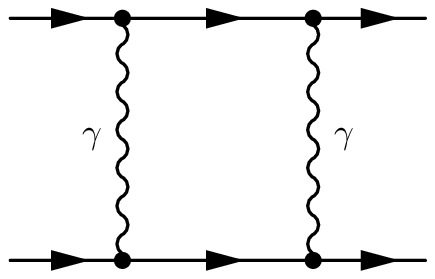}\label{FigBox1}}
        \quad
        \subfigure[]{\includegraphics[width=0.2\textwidth]{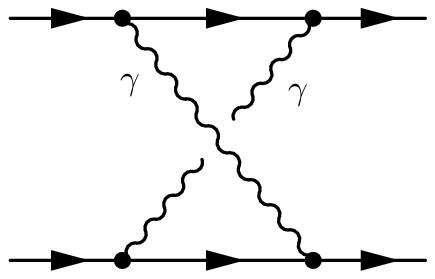}\label{FigBox2}}
    }
    \caption{Diagrams of vacuum polarization and vertex correction types.}
    \label{fig.BoxFD}
\end{figure}

Let us consider box-type diagrams
(see Fig.~\ref{fig.BoxFD}).
Using the symmetry reasons we can restrict ourself by considering only two chiral amplitudes
\eq{
    M^{++++}_B&=i\alpha^2(1-\P)\frac{1}{a b}m^{++++}, \qquad m^{++++}=M_1^{++++}+M_2^{++++}; \nn \\
    M^{+--+}_B&=-\P M^{+-+-}_B; \nn \\
    M^{+-+-}_B&=i\alpha^2\frac{1}{c_1 d_1}m^{+-+-}, \qquad m^{+-+-}=M_1^{+-+-}+M_2^{+-+-},
}
where $M_1$ and $M_2$ correspond to diagram on Fig.~\ref{FigBox1} and ~\ref{FigBox2} respectively.
Operator $\P$ interchanges the momenta of the scattered electrons
\eq{
\P F\br{p_1',p_2'}=F\br{p_2',p_1'},
}
and
\eq{
    M_1^{++++}&=\int\frac{d k}{(0)(1)(2)(q)} N_1; \qquad
    M_2^{++++}=\int\frac{d k}{(0)(1)(2')(q)} N_2; \qquad d k=\frac{d^4 k}{i\pi^2} \nn \\
    M_1^{+-+-}&=\int\frac{d k}{(0)(1)(2)(q)} M_1; \qquad
    M_2^{+-+-}=\int\frac{d k}{(0)(1)(2')(q)} M_2. \nn
}
Here the denominators are:
\ga{
    (0)=k^2-\lambda^2,\qquad (1)=(p_1-k)^2-m^2+i0; \qquad (q)=(q_1-k)^2-\lambda^2; \nn \\
    (2)=(p_2+k)^2-m^2+i0; \qquad (2')=(p_2'-k)^2-m^2+i0,
}
and the numerators are
\eq{
    N_1&=\frac{1}{2}\Sp\brs{\dd{p}_1'\gamma_\mu\br{\dd{p}_1-\dd{k}}\gamma_\nu\dd{p}_1\dd{p}_2\dd{p}_2'\gamma_\mu\br{\dd{p}_2+\dd{k}}\gamma_\nu\dd{p}_2\dd{p}_1}; \nn \\
    N_2&=\frac{1}{2}\Sp\brs{\dd{p}_1'\gamma_\mu\br{\dd{p}_1-\dd{k}}\gamma_\nu\dd{p}_1\dd{p}_2\dd{p}_2'\gamma_\nu\br{\dd{p}'_2-\dd{k}}\gamma_\mu\dd{p}_2\dd{p}_1}; \nn \\
    M_1&=\frac{1}{2}\Sp\brs{\dd{p}_1'\gamma_\mu\br{\dd{p}_1-\dd{k}}\gamma_\nu\dd{p}_1\dd{p}_2'\gamma_\mu\br{\dd{p}_2+\dd{k}}\gamma_\nu\dd{p}_2}; \nn \\
    M_2&=\frac{1}{2}\Sp\brs{\dd{p}_1'\gamma_\mu\br{\dd{p}_1-\dd{k}}\gamma_\nu\dd{p}_1\dd{p}_2'\gamma_\nu\br{\dd{p}_2'-\dd{k}}\gamma_\mu\dd{p}_2}.
}
Calculating the traces and using the table of relevant integrals, listed in Appendix~\ref{appendix.LoopIntegrals} we obtain:
\eq{
    m^{++++}(s,t,u)&=4s^3tI_{012q}-2tuI_{012'q}+4t^2(s-u)I_{01q}+4ut(s-u)I_{012'}-
    \nn\\
    &-4stl_u+4stl_t, \label{InegralsUsed} \\
    m^{+-+-}(s,t,u)&=2s(t^2-2su)I_{012q}-4u^3I_{012'q}+4t(s-u)I_{01q}+4s(s-u)I_{012}-
    \nn\\
    &-4ul_t+4ul_s.\nn
}
Inserting the explicit expressions for scalar integrals with 3 and 4 denominators (see Appendix~\ref{appendix.LoopIntegrals}) we obtain
\eq{
    m^{++++}(s,t,u)&=8s^2(l_\lambda+l_s)l_{su}+2u^2l_{tu}^2-2s^2L^2+4tsl_{tu}; \nn \\
    m^{+-+-}(s,t,u)&=\frac{8s^2}{t}l_\lambda l_{su}+4ul_{su}+2(s-u)l_{ts}^2-2s^2L^2+4tsl_{tu}-\nn\\
    &-\frac{8u^2}{t}l_sl_{us}-\frac{8u^2}{t}l_{us}l_{ts}+12(u-s)\zeta_2.\nn
}
The lowest order contribution arises from the interference term of Born amplitude with the box type amplitude
\eq{
    \Delta_B|M^{++++}|^2&=2\alpha^2(8\pi\alpha s)^2\br{\frac{1}{a b}-\frac{1}{c d}}^*\br{\frac{1}{a b}m^{++++}(s,t,u)-\frac{1}{c d}m^{++++}(s,u,t)}, \nn \\
    \Delta_B|M^{+-+-}|^2&=2\alpha^2(8\pi\alpha u)^2\br{\frac{1}{c_1 d_1}}^*\frac{1}{c_1 d_1}m^{+-+-}(s,t,u), \label{eq.Box} \\
    \Delta_B|M^{+--+}|^2&=2\alpha^2(8\pi\alpha t)^2\br{\frac{1}{c_2 d_2}}^*\frac{1}{c_2 d_2}m^{+--+}(s,t,u). \nn
}

The simple calculation of (\ref{eq.Box}) gives the total result (the contributions of vertex and soft photons (\ref{eq.SoftVirt}) are added):
\eq{
    \Delta(|M^{++++}|^2+|M^{----}|^2)&=32\pi\alpha^3
    \left\{
        \frac{8s^4}{t^2u^2}\brs{\br{l_s-1}\br{\frac{3}{2}+2l_\epsilon}+\br{l_\lambda+l_s+2l_\epsilon}\br{l_{ts}+l_{us}}}+
    \right.\nn \\
    &+\frac{8s^4}{t^2u^2}\brs{\frac{3}{2}+K_{soft}}-\frac{s}{t^2u}\brs{2u^2l^2_{tu}-2s^2\br{l_{ts}+l_{us}}^2+4tsl_{tu}}-
    \nn\\
    &-\frac{s}{u^2t}\brs{2t^2l^2_{tu}-2s^2\br{l_{ts}+l_{us}}^2+4usl_{ut}} -\nn \\
    &-\frac{8s^3}{t^2u}\brs{-\br{l_\lambda+l_s}l_{ts}+\frac{3}{2}l_{ts}-\frac{1}{2}l_{ts}^2-2+\zeta_2}-
    \nn\\
    &\left.
        -\frac{8s^3}{u^2t}\brs{-\br{l_\lambda+l_s}l_{us}+\frac{3}{2}l_{us}-\frac{1}{2}l_{us}^2-2+\zeta_2}
    \right\}.
}
We see the cancelation of terms containing the quantity $l_\lambda+l_s$.
That is the result of well known Bloch--Nordsieck theorem of infrared divergencies cancelation \cite{Bloch:1937pw}.
Similar cancelation takes place for corrections to two remaining chiral amplitudes.
As a final result we have for all contributions except vacuum polarization:
\eq{
    \Delta(|M^{++++}|^2+|M^{----}|^2)&=
    32\pi\alpha^3\left\{
        \frac{8s^4}{t^2u^2}\brs{\br{l_s-1}\br{\frac{3}{2}+2l_\epsilon}+2l_\epsilon\br{l_{ts}+l_{us}}}+
    \right.\nn \\
    &+\frac{8s^4}{t^2u^2}\brs{\frac{3}{2}+K_{soft}}-\frac{s}{t^2u}\brs{2u^2l^2_{tu}-2s^2\br{l_{ts}+l_{us}}^2+4tsl_{tu}}-
    \nn\\
    &-\frac{s}{u^2t}\brs{2t^2l^2_{tu}-2s^2\br{l_{ts}+l_{us}}^2+4usl_{ut}} -\nn \\
    &\left.
        -\frac{8s^3}{t^2u}\brs{\frac{3}{2}l_{ts}-\frac{1}{2}l_{ts}^2-2+\zeta_2}
        -\frac{8s^3}{u^2t}\brs{\frac{3}{2}l_{us}-\frac{1}{2}l_{us}^2-2+\zeta_2}
    \right\};\label{eq.AmpPPPPRC}
}
and
\eq{
    \Delta(|M^{+-+-}|^2+|M^{-+-+}|^2)&=32\pi\alpha^3
    \left\{
        \frac{8u^2}{t^2}\brs{\br{l_s-1}\br{\frac{3}{2}+2l_\epsilon}+2l_\epsilon\br{l_{ts}+l_{us}}}+
    \right.\nn \\
    &+\frac{8u^2}{t^2}\brs{\zeta_2-\frac{1}{2}-\frac{1}{2}l_{ts}^2+\frac{3}{2}l_{ts}+K_{soft}}+
    \nn\\
    &\left.
        +\frac{1}{t}\brs{4ul_{su}+12(u-s)\zeta_2-\frac{8u^2}{t}l_{us}l_{ts}+2(s-u)l^2_{ts}}
    \right\}.\label{eq.AmpPMPMRC}
}
At least
\eq{
    \Delta(|M^{+--+}|^2+|M^{-++-}|^2)=\P \Delta(|M^{+-+-}|^2+|M^{-+-+}|^2).
}

\section{Renormalization group structure of the result}
\label{SectionRenormalizationGroup}

Let us note at the moment that the corrections to chiral amplitudes squared
(\ref{eq.AmpPPPPRC}) and (\ref{eq.AmpPMPMRC}) contain the term
\eq{
    \rho=\frac{\alpha}{2\pi}\br{l_s-1}\br{\frac{3}{2}+2l_\epsilon},
}
and the rest parts of them do not contain "collinear large logarithm" $l_s$.
This term is related with $\delta$-part $P^{(1)}_\delta\br{x}$ of the kernel $P^{(1)}\br{x}$ of evolution equation \cite{Kuraev:1988xn,Jadach:2003bu} (i.e. second term in square braces)
\eq{
    P^{(1)}\br{x}&= P^{(1)}_\theta\br{x} + P^{(1)}_\delta\br{x} = \lim_{\epsilon\to 0}\brs{\theta\br{1-x-\epsilon}\frac{1+x^2}{1-x}+\delta\br{1-x}\br{\frac{3}{2}+2\ln\epsilon}},
    \label{eq.P1}
}
that is
\eq{
    \rho =
    \frac{\alpha}{2\pi}\br{l_s-1}\int\limits_0^{1}P^{(1)}_\delta\br{x}dx.
}
This fact allows us to generalize the result obtained first to include the emission of hard photons emitted in the directions of the initial and the scattered
electrons and second to take into account the contributions of higher orders of perturbation theory in leading and next to leading approximation:
\eq{
    |M^{\lambda}|^2(x_1,x_2)&=\int\limits_0^1 dy_1 D\br{y_1,l_s}\int\limits_0^1 dy_2 D\br{y_2,l_s}\frac{1}{\brm{1-\Pi_s}^2}|M^{(0)\lambda}|^2(y_1,y_2;z_1,z_2)\times\nn\\
    &\times \frac{1}{z_1}D\br{\frac{x_1}{z_1},l_s}\frac{1}{z_2}D\br{\frac{x_2}{z_2},l_s}\br{1+\frac{\alpha}{\pi}\br{K^\lambda+K_\Pi^\lambda}},
    \label{DYform}
}
where $K_\Pi^{\lambda}$ are different for different chiral states $\br{\lambda}$ and have a form:
\eq{
    K_\Pi^{++++} &=
    \frac{2}{3}\brs{
        \frac{\tilde u}{\tilde s}\br{\tilde l_{st} - \frac{5}{3}}
        +
        \frac{\tilde t}{\tilde s}\br{\tilde l_{su} - \frac{5}{3}}
    },
    \quad
    \frac{\tilde t}{\tilde s} = \frac{t}{s} \frac{z_1}{x_2},
    \quad
    \frac{\tilde u}{\tilde s} = \frac{u}{s} \frac{z_2}{x_2},
    \nn\\
    K_\Pi^{+-+-} &=
    \frac{2}{3}\br{
        - \tilde l_{st} + \frac{5}{3}
    },
    \qquad
    \tilde l_{st} = l_{st} + \ln\br{\frac{x_2}{z_1}},
    \nn\\
    K_\Pi^{+--+} &=
    \frac{2}{3}\br{
        - \tilde l_{su} + \frac{5}{3}
    },
    \qquad
    \tilde l_{su} = l_{su} + \ln\br{\frac{x_2}{z_2}},
    \nn\\
    \Pi_s &=\frac{\alpha}{3\pi}\br{l_s+\ln\br{x_1x_2}-\frac{5}{3}}.
}
The Structure functions $D\br{x,l_s}$ in (\ref{DYform}) obey the QED renormalization (evolution) equations and can be put in form of the series \cite{Kuraev:1988xn,Jadach:2003bu}
\eq{
D\br{x,l} &= \delta\br{1-x}+\frac{\alpha}{2\pi}\br{l-1}P^{(1)}(x)+\frac{1}{2!}\br{\frac{\alpha}{2\pi}\br{l-1}}^2P^{(2)}\br{x}+... \nn
}
where first order kernel $P^{(1)}(x)$ is presented in (\ref{eq.P1}) and the higher order kernels $P^{(n)}(x)$, $n\geq2$ can be
obtained by the following recursive relation:
\eq{
P^{(n)}\br{x}&= \int\limits_x^1\frac{d y}{y}P^{(1)}(y)P^{n-1}\br{\frac{x}{y}}.
}
The $K$-factor in (\ref{DYform}) represents thus the non-leading terms of correction and usually of order of unity.
It can be obtained if one put $\rho=0$ in the above chiral amplitudes (\ref{eq.AmpPPPPRC}) and (\ref{eq.AmpPMPMRC})
(except vacuum polarization) and extract the general Born
amplitude squared.
The quantity $|M^{(0)}|^2(x_1,x_2;z_1,z_2)$ represents the Born amplitude square in shifted kinematics,
where $x_1$, $x_2$, $z_1$, $z_2$ are the energy fractions of initial and final particles.
The energy fractions $z_1$, $z_2$ can be found using the conservation laws:
\eq{
x_1 + x_2 = z_1 + z_2,
\qquad
x_1 - x_2 = z_1 C_1 + z_2 C_2,
\qquad
0 = z_1 S_1 + z_2 S_2,
}
where $C_{1,2} = \cos{\theta_{1,2}}$, $S_{1,2} = \sin{\theta_{1,2}}$, $\theta_{1,2} = \widehat{\vv{p_1},\vv{p_{1,2}'}}$, which
gives:
\eq{
    z_1 = \frac{2x_1 x_2}{x_1\br{1-C_1} + x_2\br{1+C_1}},
    \quad
    z_2 = \frac{x_1^2\br{1-C_1} + x_2^2\br{1+C_1}}{x_1\br{1-C_1} + x_2\br{1+C_1}}.
}

\section{Conclusion}
\label{SectionConclusion}

In this paper the compact expressions for chiral amplitudes of  \Moller  scattering
free from infrared divergences are calculated  on Born and one-loop QED level.
The separate chiral amplitudes contribution to
the cross section is in agreement with renormalization group predictions.
When inferring this formulae we systematically omitted terms of order
$    {\cal O} \br{ \br{{\alpha}/{\pi}}^2 {m^2}/{s}  }$.
The explicit form of $K$-factors provides the accuracy of this results to
leading and next-to-leading logarithm approximation. This results into accuracy
up to one permille.

In conclusion we note that the unpolarized case was considered in leading and next-to-leading approximation in
\cite{Arbuzov:2011zzb}, where the cross section of \Moller scattering with additional hard photon emission
is presented as well.

\acknowledgements

The authors acknowledge to RFBR 11-02-00112-a and Belorussian 2011 year grants for financial support.
One of us (Yu.~M.~B.) also acknowledges JINR grant for Young Scientists of 2011 for support.


\appendix

\section{Table of integrals}
\label{appendix.LoopIntegrals}

In this Appendix we present the set of vector and scalar loop momentum integrals
which we used in formulae (\ref{InegralsUsed}) and below it.

Vector integrals are expressed in terms of scalar ones:
\eq{
    &\int\frac{d^4 k}{i \pi^2}
    \frac{k^\mu}{\br{0}\br{1}\br{2}\br{q}}
    = a\br{p_1-p_2}^\mu + b q^\mu,
    \qquad
    q = p_1 - p_1',
    \\
    &a = \frac{1}{2u}\br{A+B},
    \qquad
    b = -\frac{1}{2 t u}\br{A\br{t-u}-s B},
    \nn\\
    &A = I_{012} - I_{02q},
    \qquad
    B = A - t I_{012q}.
    \nn\\
    &\int\frac{d^4 k}{i \pi^2}
    \frac{k^\mu}{\br{1}\br{2}\br{q}}
    = q^\mu\br{I_{12q} - 2\frac{l_s}{s}} + \Delta^\mu \frac{l_s}{s},
    \qquad
    \Delta = p_1 - p_2,
    \nn\\
    &\int\frac{d^4 k}{i \pi^2}
    \frac{k^\mu}{\br{0}\br{1}\br{2}}
    = \frac{l_s}{s}\Delta^\mu,
    \nn\\
    &\int\frac{d^4 k}{i \pi^2}
    \frac{k^\mu}{\br{0}\br{1}\br{q}}
    = p_1^\mu \br{I_{01q} - 2\frac{l_t}{t}} + \frac{1}{t} l_t q^\mu,
    \nn\\
    &\int\frac{d^4 k}{i \pi^2}
    \frac{k^\mu}{\br{0}\br{2}\br{q}}
    = p_2^\mu \br{-I_{02q} + 2\frac{l_t}{t}} + \frac{1}{t} l_t q^\mu,
    \nn\\
    &\int\frac{d^4 k}{i \pi^2}
    \frac{k^\mu}{\br{0}\br{1}\br{2'}\br{q}}
    = \alpha\br{p_1+p_2'}^\mu + \beta q^\mu,
    \nn\\
    &\alpha = \frac{1}{2s}\br{A'+B'},
    \qquad
    \beta = -\frac{1}{2 s t}\br{A'\br{t-s}-u B'},
    \nn\\
    &A' = I_{12'q} - I_{02'q},
    \qquad
    B' = A' - t I_{012'q}.
    \nn\\
    &\int\frac{d^4 k}{i \pi^2}
    \frac{k^\mu}{\br{1}\br{2'}\br{q}}
    = q^\mu\br{I_{12'q} - 2\frac{l_u}{u}} + \br{p_1+p_2'}^\mu \frac{l_u}{u},
    \nn\\
    &\int\frac{d^4 k}{i \pi^2}
    \frac{k^\mu}{\br{0}\br{1}\br{2'}}
    = \frac{l_u}{u}\br{p_1+p_2'}^\mu,
    \nn\\
    &\int\frac{d^4 k}{i \pi^2}
    \frac{k^\mu}{\br{0}\br{2'}\br{q}}
    = p_2'^\mu \br{I_{02'q} - 2\frac{l_t}{t}} + \frac{1}{t} l_t q^\mu.
    \nn
}
The scalar integrals which encounter in expressions above have a form:
\eq{
    \Re I_{012q} &=
    \Re\int\frac{d^4 k}{i \pi^2}
    \frac{1}{\br{0}\br{1}\br{2}\br{q}}
    =
    \frac{2}{s t} l_s \br{l_t+l_\lambda},
    \nn\\
    \Re I_{012} &= \Re I_{12q} =
    \Re\int\frac{d^4 k}{i \pi^2}
    \frac{1}{\br{0}\br{1}\br{2}}
    =
    \frac{1}{2s} \br{l_s^2 - 8\zeta_2 + 2\ l_s l_\lambda},
    \nn\\
    \Re I_{01q} &= \Re I_{02q} = \Re I_{02'q} =
    \Re\int\frac{d^4 k}{i \pi^2}
    \frac{1}{\br{0}\br{1}\br{q}}
    =
    \frac{1}{2t} \br{l_t^2 + 2\zeta_2},
    \nn\\
    \Re I_{012'q} &=
    \Re\int\frac{d^4 k}{i \pi^2}
    \frac{1}{\br{0}\br{1}\br{2'}\br{q}}
    =
    \frac{2}{t u} l_u \br{l_t+l_\lambda},
    \nn\\
    \Re I_{012'} &= \Re I_{12'q} =
    \Re\int\frac{d^4 k}{i \pi^2}
    \frac{1}{\br{0}\br{1}\br{2'}}
    =
    \frac{1}{2u} \br{l_u^2 - 8\zeta_2 + 2\ l_u l_\lambda}.
    \nn
}

\end{document}